\documentstyle[manuscript,aps]{revtex}
\begin{document}
\draft
\title{Scaling Theory and Numerical Simulations of
Aerogel Sintering}
\author{R\'emi Jullien, Nathalie Olivi-Tran, Anwar Hasmy, Thierry Woignier,
Jean Phalippou, Daniel Bourret and Robert Semp\'er\'e}
\address{Laboratoire de Science des Mat\'eriaux Vitreux,
Universit\'e Montpellier II, Place Eug\`ene Bataillon,
34095 Montpellier, France}
\date{\today}
\maketitle
\begin{abstract}
A simple scaling theory for the sintering of fractal aerogels is presented. The
densification at small scales is described by an increase of the lower cut-off
length $a$  accompanied by a decrease of the upper cut-off
length $\xi$, in order to conserve the total mass of the system. Scaling laws
are derived which predict how $a$, $\xi$ and the specific pore surface area
$\Sigma$ should depend on the density $\rho$.
Following the general ideas of the theory, numerical simulations of
sintering are proposed starting from computer simulations of
aerogel  structure based on a diffusion-limited cluster-cluster aggregation
gelling process. The numerical results for $a$, $\xi$ and $\Sigma$ as a
function of $\rho$ are discussed according to the initial aerogel density. The
scaling theory is only fully recovered in the limit of very low density where
the original values of $a$ and $\xi$ are well separated. These numerical
results are compared with experiments on partially densified aerogels.
\end{abstract}


\section{Introduction}

Sintering of silica aerogels is a process in which the original material is
heated
at a temperature smaller than the melting temperature
of the corresponding silica crystal. It results in
a strengthening of the internal structure and a gradual elimination of the
pores
accompanied by a general densification  and shrinkage of the whole
sample.
Such process allows the design of intermediate materials, hereafter called PDA
(partially densified aerogels) of increasing density
up to full dense silica glass. Several theories have been introduced to explain
the shrinkage due to sintering \cite{1,2,3,4,5}.  It is now quite well
established that
the internal mechanism for sintering of aerogels is due to local viscous
flows of matter. For the application to glasses and aerogels, the most
convincing theoretical approaches are due to Scherer \cite{4,5}. Here, we
present a
 scaling approach, which can be considered as a generalization of
Scherer's type of calculations to fractal matter. Since a short account of
the theory has already been published elsewhere \cite{6}, we present here a
slightly
different (simpler) derivation of the scaling laws (part 2A).
Based on these theoretical ideas, we present  two
recently developed numerical procedures able to simulate the sintering of
realistic aerogel structures (part 2B). Then, after presenting the numerical
results (part 3), we compare them with experiments (part 4) and we conclude
(part 5).

\section{Constraints on theory}
\subsection{Scaling theory}

It is now well established (see
for example the recent interpretation of low angle neutron scattering
experiments \cite{7}) that silica aerogels are made of connected fractal
aggregates. Fractal scaling occurs in a range of lengths extending from a
lower cut-off $a$, which is the mean diameter of the silica particles
constituting the aggregates, up to an upper cut-off $\xi$ which is the mean
diameter of the aggregates, or, equivalently, the mean center-to-center
distance between two neighboring connected aggregates. Since it is admitted
that, in aerogels, sintering results from local viscous flow of matter, we will
assume that only the shortest lengths are concerned and therefore only
the lower cut-off $a$ is affected, without changing neither the fractal
dimension of the aggregates nor their mutual arrangement.  After a
first stage where the individual shapes of silica particles are only
slightly modified, while bridges are built between them, one reaches a
``scaling'' regime where individual particles can no longer be distinguished.
In this regime, we will assume that there still exist a lower cut-off
$a$, which can then be defined as the mean thickness of the aggregates arms.
Since the sintering process tends to lower the internal surface pore area
by transfering matter from large curvature  to small curvature regions and
therefore reinforcing the thinner arms, there results  a gradual increase of
$a$. But, correlatively, to insure the mass conservation, there should be a
general shrinkage of the material, characterized by a decrease of $\xi$.
Figure 1 gives a two-dimensional sketch of how an aggregate looks like at two
different stages of sintering.

Since we assume that the overall inter-aggregate structure is
conserved, the shrinkage at the scale of an aggregate is the same as at the
macroscopic scale. Therefore, if $\rho$ is the mean density of the material,
the following ``trivial'' scaling should hold:
\begin{eqnarray}
\xi \sim \rho^{-{1\over
3}}
\label{E1}
\end{eqnarray}
Then, using the results of the fractal geometry for simple
scale-invariant structures \cite{8}, the
minimum number $N$ of balls of diameter $a$ necessary to cover the total mass
of a fractal aggregate of diameter $\xi$ (the requirement being that each
point of silica matter should be inside at least one ball), is given by:
\begin{eqnarray}
N\sim ({\xi\over a})^D
\label{E2}
\end{eqnarray}
 Very often, this formula appears in the literature with only
one parameter ($a$ or $\xi$) because the other one is assumed to stay
constant. Given $a$, the relation $N\sim \xi^D$ corresponds to the
``mass-size'' relation. Given $\xi$, the relation $N\sim a^{-D}$ corresponds
to the Haussdorf-Besicovitch definition of the fractal dimension (for simple
self-similar fractals all the definitions of the fractal dimension lead to the
same value \cite{8}). But here, we need to keep the two parameters together
since
they both vary. Using this formula, the bulk density can be expressed as a
function of $a$ and $\xi$:
\begin{eqnarray}
\rho \sim {N a^3\over \xi^3} \sim ({a\over
\xi})^{3-D}
\label{E3}
\end{eqnarray}
Then, combining with equation (1), we immediately get:
\begin{eqnarray}
a \sim \rho^{D/3(3-D)}
\label{E4}
\end{eqnarray}
Even if the present derivation of equation (4) is more straightforward than
the one already given  \cite{6}, it  implicitely uses the same simple
symetries (fractal scaling and mass conservation).

As soon as $a$ and $\xi$ are known, any other structural property, which is
related to these parameters, can be calculated. Several examples have been
previously provided \cite{6}  but here we will focus on the specific pore
surface area $\Sigma$ which is the area of the whole silica-air interface
counted per unit of silica mass. Following the above ball-covering reasoning,
the total interface for one aggregate, is of order $Na^2$  while its
mass is of order $Na^3$, therefore one has:
\begin{eqnarray}
\Sigma \sim {Na^2\over Na^3}\sim {1\over a} \sim
\rho^{-D/3(3-D)}
\label{E5}
\end{eqnarray}

A key point of our reasoning is that $a$ remains a well defined cut-off. In
other words
the
constitutive aggregates can be modelized by {\it mass fractals} downs to length
$a$. Therefore
to satisfy (1) and (3) simultaneously, $a$ should increase if $\xi$ decreases
and reciprocally. Such reasoning fails if the interface might be considered as
a
{\it surface fractal}; then one could imagine to smooth the surface without
changing $\xi$. This may happen for other kind of materials and, here,
for the last stages of sintering where the
scaling theory doesnot work.

Before pursuing, it might interesting to compare our scaling results with
previous
Scherer's approach of sintering \cite{4,5}. In its first
calculation \cite{4}, Scherer was considering
a regular cubic array whose bonds were made of cylinders of length $\xi$ and
diameter $a$. Neglecting surface deformations near the cylinders connections
and simulating the sintering as an increase of $a/\xi$, he was able to
calculate analytically the density as a function of $a/\xi$, at least for
$a<\xi$:
\begin{eqnarray}
{\rho\over\rho_S} = 3\pi ({a\over\xi})^2 - 8\sqrt{2}
({a\over\xi})^3
\label{E6}
\end{eqnarray}
where $\rho_S$ is the density of silica.
This formula can be usefully compared with our
formula (3). Since Scherer is considering the peculiar case of $D=1$ fractals
his first term is the same as ours, but, in his simple geometry, he is
able to perform the full analytical calculation and he gets a second term
which,
in our language, appears to be a correction to scaling since it is negligeable
compared to the first term when $\xi /a$ is large.

\subsection {Numerical simulations}

Since it has been shown that the diffusion-limited cluster-cluster aggregation
model \cite{9,10} provides a quite realistic modelization of an aerogel
structure \cite{7,11}, in this section we intend to use such a model
to develop numerical simulations of aerogel sintering.
This is still an approximate treatment of
sintering since we will  use some naive coarse-graining procedures, such
as that used in real-space renormalization group methods \cite{12}, to describe
the
smoothing at the shortest scale, but, since we are now considering a
realistic structure, we hope to get some informations on corrections to
scaling.

We have developed two numerical methods. The first one works on a cubic
lattice and considers cubic particles of edge length $a_0$ (volume $v_0 =
a_0^3$, mass $m_0 = \rho_S a_0^3$). The second one
works off-lattice and considers spherical particles of diameter $a_0$
(volume $v_0 = {\pi\over 6}a_0^3$, mass $m_0 = {\pi\over 6}\rho_S a_0^3$). In
both cases the particles are inside a cubic box of edge length $La_0$ and
their number $N_0$ is such that the volume fraction $c_0$ is set to a
desired value $c_0 = N_0v_0/L^3a_0^3$. Therefore, except $L$ which should be
chosen as large as possible, the only parameter of the model is $c_0$ which
is directly related to the initial aerogel density $\rho_0 = c_0\rho_S$.

To build the original (non-sintered) aerogel structure, the
particles are first randomly disposed (i. e. put on randomly chosen sites,
avoiding multiple occupancy,  on lattice, or sequentially centered at random
points, avoiding overlaps, off-lattice) in the box. Then, these particles are
allowed to undergo a brownian diffusive motion and they irreversibly stick
when come on contact. Aggregates of particles are also able to diffuse
together with the individual particles and to stick to particles or to other
aggregates. In this diffusive motion, the diffusion constant of the
aggregates is considered to vary as the inverse of their radius of gyration
and periodic boundary conditions are assumed at the box edges \cite{9,10,11}.
When the
concentration $c_0$ is sufficiently large (larger than a threshold value
$c_g$ which tends to zero for infinite box size \cite{7}), the final structure
is a gelling network which extends from edge to edge in the box and which
can be described as a loose random packing of connected fractal aggregates,
of fractal dimension $D\simeq 1.8$, whose mean size $\xi_0$ decreases as
$c_0$ increases.

The modelization of the sintering process differs if one considers the
on-lattice or the off-lattice version of the model.

\subsubsection {On lattice}

Here we have used a discrete blocking method (based on ``box counting'' ideas
\cite{8}) which proceeds step by
step. It implies that the value of $L$ is set to some power
of 2, $L = 2^n$. At step $p$ of sintering, the original cubic lattice is
replaced by a ``super-lattice'' of parameter length $2^pa_0$, each
super-cell of the new lattice containing $2^{3p}$ cells of the original
lattice. Then a super-cell is considered as occupied if it contains at least
one particle of the original structure. The number $N_p$ of occupied
super-cells is computed. Then the sintered structure is
obtained by considering the structure made of the $N_p$ occupied
super-cells and by applying an adequate length contraction $\beta_p$ in order
to conserve the total mass. The volumic fraction $c_p$ of the super-structure
being:
\begin{eqnarray}
c_p = {N_p(2^pa_0)^3\over (La_0)^3} = 2^{3p}{N_p\over
N_0}c_0
\label{E7}
\end{eqnarray}
the contraction factor is given by:
\begin{eqnarray}
\beta_p = ({c_p\over c_0})^{1\over 3} = 2^p ({N_p\over N_0})^{1\over
3}
\label{E8}
\end{eqnarray}
and the actual value of the cut-offs $a_p$ and $\xi_p$ at step $p$ are given
by:
\begin{eqnarray}
a_p = 2^p{a_0\over\beta_p}= ({N_p\over N_0})^{1\over 3}a_0
\label{E9}
\end{eqnarray}
\begin{eqnarray}
\xi_p = {\xi_0\over\beta_p}
\label{E10}
\end{eqnarray}
As expected when combining formulae (8) and (10) one recovers that  the
``trivial'' scaling law (1) on $\xi$ is automatically
satisfied, but the one on $a$ deserves to be tested by the numerical
calculations. On figure 2 (a) we provide a typical example where a section of
the box is shown at three different steps of sintering.

To determine the specific pore surface area, we numerically determine the
number $S_p$ of square interfaces separating occupied supercells nearest
neighbor to a non-occupied one (taking care of the periodic boundary
conditions) and we calculate $\Sigma$ by:
\begin{eqnarray}
\Sigma_p = {1\over \rho_S}{S_p
a_p^2\over N_p a_p^3} = {1\over a_0\rho_S} {S_p\over N_p^{2\over
3} N_0^{1\over 3}}
\label{E11}
\end{eqnarray}
In practice, for a given $c_0$ and at each step $p$, the quantities $N_p$ and
$S_p$, calculated numerically, have been averaged over several realizations of
the initial configuration.

Note that we could have removed the
restriction to supercells of
edge lengths $2^pa_0$, working instead with edge lengths $\ell a_0$,
$\ell$ being
any integer. Imagining such an extension of the method, let us call $\ell_f$
the
value of $\ell$ above which all the supercells become occupied and define
$\xi_0$
as being $\ell_fa_0$. With this definition $\xi_0$ is not, rigourously
speaking, the mean aggregate diameter but, rather, it is such that
$\xi_0-a_0$ represents the edge length of the largest cubic hole in the
structure.
Using then the fact that, at the end of the sintering process, one has $\xi_f =
a_f$, one gets:
\begin{eqnarray}
a_f = \xi_0 c_0^{-{1\over 3}}
\label{E12}
\end{eqnarray}
In practice we have written a separate code  to evaluate $\xi_0$ by determining
$\ell_f$. Therefore using (12), one can get a numerical estimate of $a_f$.

\subsubsection {Off lattice}

Here we have used a dressing method which, compared to the preceeding method,
has the advantage to be continuous. We first consider a $s$-dependent
dressed structure, where $s$ is a continuous variable, in which each initial
sphere of diameter $a_0$ is replaced by a sphere of the same center but of
larger diameter $a_d(s)$, given by:
\begin{eqnarray}
a_d(s) = a_0(1+s)
\label{E13}
\end{eqnarray}
Then the total volume $V_d(s)$ located inside the overlapping spheres is
numerically calculated (avoiding multiple counting of overlaps). The volumic
fraction of the dressed structure is given by:
\begin{eqnarray}
c(s) = {V_d(s)\over (La_0)^3}
\label{E14}
\end{eqnarray}
Here also, to insure mass conservation, the sintered structure is obtained from
the dressed
structure after applying an adequate length contraction $\beta(s)$ given by:
\begin{eqnarray}
\beta(s) = ({c(s)\over
c_0})^{1\over 3}= ({6V_d(s)\over N\pi a_0^3})^{1\over 3}
\label{E15}
\end{eqnarray}
Then the actual diameter  $a(s)$ of the spheres of the sintered  structure and
the correlation length $\xi(s)$ are given by:
\begin{eqnarray}
a(s) ={a_d(s)\over\beta(s)}= a_0{1+s\over \beta(s)}
\label{E16}
\end{eqnarray}
\begin{eqnarray}
\xi(s) = {\xi_0\over\beta(s)}
\label{E17}
\end{eqnarray}
Here again the scaling law on $\xi(s)$ is automatically verified while $a(s)$
needs a numerical calculation. The value for $\xi_0$ is estimated from $a_f$,
the limiting value of $a(s)$ when $c(s)$ tends to one, by inverting
formula (12). Here $\xi_0-a_0$ represents the diameter of the largest
spherical hole. On figure 2(b) we provide a typical example
where a section of the box is shown at three different steps of sintering.

To determine the specific pore surface area $\Sigma(s)$, we observe that the
total surface $S_d(s)$ of the dressed structure is simply related to the
derivative of the function $V_d(s)$:
\begin{eqnarray}
S_d(s) = 2{dV_d\over da_d} = {2\over a_0}{dV_d(s)\over ds}
\label{E18}
\end{eqnarray}
Then, dividing by the total mass and correcting by the adequate scaling factor,
we get the following expression $\Sigma(s)$:
\begin{eqnarray}
\Sigma(s) = {2\over a_0\rho_S}{1\over V_d(s)}{dV_d\over ds}\beta(s) =
{2\over a_0\rho_S}{d\log c(s)\over ds}\beta(s)
\label{E19}
\end{eqnarray}
In practice, for a given $c_0$, we have numerically determined the whole curve
$c(s)$ versus $s$ as well as its derivative and we have calculated,
$a(s)$ and $\Sigma(s)$ using formulae  (16) and (19).  Here also, for each
$c_0$ value, the results
have been averaged over several realizations of the intial configuration.

\section{Results}

In both the on-lattice and off-lattice cases, we have made calculations with
$L=64$, we have considered  initial concentrations $c_0 = 0.01, 0.02, 0.05$
and 0.1 and the results for $c$, $a$ and $\Sigma$ have been averaged over 20
independent realizations of the initial configuration. Figure 3
gives the results for $a/a_0$ and $\xi/a_0$ as a function of $c$ (log-log
plot) for the different $c_0$ values. Cases (a) and (b) correspond to
on-lattice (symbols) and
off-lattice (continuous lines) results, respectively. The slope indicated is
the
theoretical slope $D/3(3-D)=0.5$ expected from the scaling theory with
$D=1.8$. In principle, the scaling theory should be recovered in the vanishing
concentration limit $c_0\rightarrow 0$ where $a_0$ and $\xi_0$ are well
separated. Even if the results for $a$ versus $c$ are more and more linear,
with an apparent slope close to 0.5,  when $c_0$ decreases, in both cases we
observe, down to $c_0=0.02$, some corrections to scaling both at low $c$ and
large $c$.

The low-$c$ corrections to scaling are different in the two simulations.
When performing a low-$s$ expansion in the off-lattice formulae, it can be
shown that $a$ in this case should behave quadratically when $c$ tends to
$c_0$:
\begin{eqnarray}
({a-a_0\over a_0})_{\hbox{off}}\sim ({c-c_0\over c_0})^2
\label{E20}
\end{eqnarray}
This singular behavior is due to the peculiar non sintered structure made of
perfectly tangent spheres. In all other situations, with non zero areas between
connected particles, as it is in the on-lattice case, the low-$c$ behavior
should be linear:
\begin{eqnarray}
({a-a_0\over a_0})_{\hbox{on}}\sim {c-c_0\over c_0}
\label{E21}
\end{eqnarray}
Therefore, in the off-lattice case, there is an initial regime during which
finite areas are grown between connected particles. During this regime, for
the same increase of $\rho$, $a$ increases less than in the on-lattice case.
After this regime, the evolution of $a$ becomes similar to the one of the
on-lattice case, but for a larger initial concentration. This is shown in
figure 4 (a) where we have reported the off-lattice results for $c_0 =
0.01$(dashed curve),
which, except at low concentrations, are very close to the on-lattice results
for $c_0 = 0.02$.
Note that, apart from low-$c$ corrections
to scaling, the lattice structure
built with spherical particles of diameter $a_0$, instead of cubes, which has a
concentration smaller ($c'_0=(\pi/6)c_0$), behaves quite similarly during
sintering. This is reasonable, since, starting from the same spheres it is
known that the on-lattice and off-lattice diffusion limited cluster-cluster
aggregation processes lead to the same large distance correlations \cite{13}.

At large $c$ ($c$ close to 1) the correction to scaling comes from the fact
that,
in this limit, it remains only a few holes in the structure and, obviously, one
can no longer speak of a structure made of connected fractals. The volume
fraction of the last
hole being $1-c$ and its radius being proportional to $a_f-a$, where $a_f$ is
the limiting value of $a$,  one should have:
\begin{eqnarray}
a_f - a \sim (1-c)^{1\over 3}
\label{E22}
\end{eqnarray}

This behavior is well verified by both on-lattice and off-lattice numerical
results.

Concerning the specific pore surface area, we have  calculated the
dimension-less quantity $a_0\rho_0\Sigma$. It is worth noticing that we get
different results already for $c=c_0$. In the off-lattice case, we recover
the trivial result:
\begin{eqnarray}
(a_0\rho_0\Sigma_0)_{\hbox{off}} = 6
\label{E23}
\end{eqnarray}
which means that the structure is made of non-overlapping spheres. In the
on-lattice case, we get a lower value since we do not count the interfaces
between neighboring particles. If we neglect the contributions of loops,
it can be shown that the 6 should be replaced by $4-2/N_0$ (since there
should be $N_0-1$ bonds). But, due to the presence of loops, the actual
number is smaller, close to 3.5 and decreases when $c_0$ increases.
The numerical results for $\Sigma$ as a function of $c$ are given in figures 4
(a) and (b) as a log-log plot of $\Sigma\over\Sigma_0$ versus $c$ for the
different $c_0$ values. In both cases the slope -0.5 expected from the
scaling theory is better verified than for $a$. Here also the $c_0$ for the
off-lattice case is close to the curve for $c'_0 =c_0/2$ for the
on-lattice case (but not as close as in the case of $a$). When $c$ tends to 1,
we recover that $\Sigma$ tends to zero, which is an
improvement compared to the toy model.   Following the above reasoning for $c$
close to 1, one should get:
\begin{eqnarray}
\Sigma\sim{1\over a_f-a}\sim (1-c)^{-{1\over 3}}
\label{E24}
\end{eqnarray}
a behavior quite well observed in our off-lattice results.

Another result of our numerical calculations is the estimate of $\xi_0$,
which is here related to the largest spherical hole in the non sintered
aerogel. In table I, we compare our  off-lattice results with the alternative
estimates  (here called $\xi'_0$) obtained from the location of the minimum of
the pair correlation function $g(r)$ \cite{7}. It is reasonable that
these  estimates are close and roughly proportional to each other.

\section{Discussion}

In this section we would like to discuss our numerical results at the
light of several experimental results on partially densified aerogels
(PDA).  Let us discuss first some previous low angle neutron diffraction
experiments \cite{14}.
In figure 6, we show the scattering intensity $I(q)$ curves for a series of
PDA of increasing
densities made by sintering a ``neutral'' aerogel, hereafter called N46. These
experimental results are in good qualitative agreement with the theory. The
slope of the fractal linear regime (here corresponding to $D\sim 2.3$) does
not depend on the density, except in the last steps of sintering (where
anyway the slope can no more be interpreted as a fractal dimension), but its
$q$ extension is gradually reduced as $\rho$ increases. The arrows show how the
parameter $a$ and $\xi$ have been previously estimated, i. e. by assuming that
$a^{-1}$ and $\xi^{-1}$ correspond to the lower and uper departures from
linearity. At the light of the recent interpretation of the $I(q)$ curves for
non-sintered aerogels \cite{7}, we now know that such method is approximate and
cannot give the right absolute values for both $a$ and $\xi$. In particular,
since $\xi$ should be related to the position of the low-$q$ maximum, not
visible on the figure, its actual value should be considerably larger than the
above estimate. However, since the estimation has been done with exactly the
same procedure for all densities, we can think that the relative values
$a/a_0$ and $\xi/\xi_0$ are relevant. The same method has been applied
for another neutral sample,  N26, of lower initial density but of similar
fractal dimension and on a basic sample, B46 of lower fractal dimension $D\sim
1.8$, for which it is believed that  the DLCA model applies \cite{7}.

Figure 6 (a) gives the results obtained with the basic sample as a
log-log plot of both $a/ a_0$ and $\xi/ a_0$ versus $\rho/\rho_S$ and the
best fits of $a/a_0$ with both calculations.
The continuous curve is the result of the off-lattice calculation with $c_0
= 0.04$ and the circles correspond to the on-lattice calculation with $c_0 =
0.08$. Since  the actual  density
of
the non-sintered sample  is about 0.2g/cm$^3$, the experimental $c_0$ value
should be about 0.09, closer to the value taken for the on-lattice fit.
The fact that the on-lattice model works better is certainly due to the
presence of finite areas between connected particles already in the
non-sintered material. Such ``pre-sintering'' might have occured during the
supercooling process used to obtain the aerogel from the gel by extracting the
solvant. The data for $a$ are quite well fitted except for the point of
largest density. Anyway, since this point corresponds to a very large density,
the method of determining $a$ from the $I(q)$ data becomes highly
questionable. Note that in the last stages of sintering, one cannot interpret
$I(q)$ by the fractal theory and one needs another numerical calculations of
$I(q)$ (such as the scattering by a collection of random polydisperse holes of
radii $a_f-a$ at averaged mutual distance $\xi$) to be able to interpret the
scattering curves. Moreover the numerical estimates of $\xi$ (except the last
point) exhibit a nice $-1/3$ slope providing a strong support to the main
hypothesis of our theory. However the estimates are smaller (by a factor
almost 5) than the theoretical values, as expected if one remembers that the
theoretical $\xi_0$ value is related to the location of the maximum of the
$I(q)$ curve.

Figure 6 (b) gives the experimental results for two neutral samples with
different initial densities. Here
we cannot try to quantitatively compare with our numerical results since the
fractal dimension is larger. It is known that  the formation of neutral
aerogels cannot be explained by the DLCA model. They have a tenuous
and flexible polymeric structure and therefore deformations and
restructuring cannot be avoided during aggregation. However the  effective
slope of the $a$ versus $c$ curves is larger than in the basic case  in
perfect aggreement with the scaling theory, as already mentionned in \cite{6}.
Assuming a fractal dimension of 2.3, the theoretical slope $D/3(3-D)$ should
be slightly larger than 1. Assuming that the corrections to scaling enters
similarly than in the basic case, the experimental slope is smaller as
seen in the figure. Moreover, here again, the results for $\xi$ exhibit a nice
$-1/3$ slope.

\section{Conclusion}

In this contribution we have presented a scaling model for the sintering
of aerogels, we have etayed this approach by some numerical calculations and
we have compared the numerical results with some experiments. The numerical
calculations have shown that corrections to scaling are not negligeable, in
particular at the last stages of sintering where the fractal scaling process
is no more valid (the notion of cut-off fails) and where the sintering process
rather involves a gradual elimination of residual holes. In the first stages of
sintering the corrections to scaling have a larger role in the $c$-dependence
of the lower cut-off $a$ than in that of the specific pore surface area. The
effective slope of the numerical $\log a$ versus $\log c$ curve is slightly
smaller than the one expected from the scaling theory. These theoretical
results are in quantitatively and qualitatively good agreement with
experimental results on basic and neutral aerogel, respectively. A quantitative
agreement in the case of neutral aerogels requires a specific modelization of
these materials which have a larger fractal dimension than basic aerogels.
Nevertheless, the numerical calculations presented here are still using crude
modelizations of what might be the actual sintering process. Even if the two
numerical methods presented here give almost the same results, in both cases,
at each step of sintering, the internal surface of the pores is not a minimum
area surface as it should be. In particular, in the last stages of sintering,
the remaining holes are not spherical. A more sophisticated numerical approach
would be to consider a minimum area surface at each step of restructuring, its
area decreasing gradually during the sintering process.  A Monte Carlo
modelization of such process is under progress. Moreover, we are presently
considering the off-lattice numerical approach as a basis to simulate the gas
transport properties of partially densified aerogels in order to interpret
some new experiments on the permeability of gas \cite{15}.

We would like to ackowledge discussions with Marie Foret, Jacques Pelous,
Ren\'e' Vacher and Peter Pfeifer. One  of us (A. H.) would like to acknowledge
support from CONICIT
(Venezuela).

     \null\vfill\eject
\centerline{TABLES}
\def\tvi{\vrule height 12pt depth 5pt width 0pt}
\def\tv{\tvi\vrule}

$$\vbox{\offinterlineskip\halign
{ #&\tv#&\quad #&\tv#&\quad #&\tv#&\quad #&\tv#&\quad #\cr
\tvi  $c_0$   && $a_f$ && $\xi_0$  && $\xi'_0$ \cr   \noalign{\hrule}  \tvi
       0.01   && 6.2 && 23   &&  28.8   \cr
       0.02   &&  4.6 && 13.5   &&  16.9  \cr
       0.05   &&  3.2 && 6.1  &&  8.7  \cr
       0.077   &&  2.5 && 4.0   &&  5.9  \cr
       0.1   &&  2.3 && 3.0   &&  5.0  \cr
  }}$$

{TABLE I}. For each concentration $c_0$ considered in the
off-lattice simulations, we have reported the value of $a_f$, the
corresponding value of $\xi_0$ and the location $\xi'_0$ of the minimum of the
pair correlation function $g(r)$.

\begin{figure}
\caption{Two dimensional sketch of an individual aggregate  at two different
stages of sintering.}
\end{figure}
\begin{figure}
\caption{Section of the box for the original configuration and at two
stages of sintering. (a) and
(b) correspond to  the on-lattice (with $L=32$) and off-lattice (with $L=40$)
method respectively.}
\end{figure}
\begin{figure}
\caption{Log-log plot of $a/a_0$ and $\xi/a_0$ versus $c =
\rho/\rho_S$ for $L=64$ and different initial concentrations $c_0$. Cases (a)
and (b) correspond to  the on-lattice  and off-lattice
method respectively. In case (a) open circles, black squares, open diamonds and
black triangles corresopond to $c_0 = 0.01, 0.02, 0.05$ and 0.1, respectively.
The solid lines are the results for $\xi/a_0$. The dashed curve shown in (a)
corresponds to the off-lattice results with $c_0 = 0.01$.}
\end{figure}
\begin{figure}
\caption{Log-log plot of $\Sigma/\Sigma_0$ versus $c = \rho/\rho_S$ for
$L=64$ and different initial concentrations $c_0$.
Cases (a)
and (b) correspond to  the on-lattice  and off-lattice
method respectively. In case (a) open circles, black squares, open diamonds and
black triangles corresopond to $c_0 = 0.01, 0.02, 0.05$ and 0.1, respectively.
The dashed curve shown in (a) corresponds to the off-lattice results with
$c_0 = 0.01$.}
\end{figure}
\begin{figure}
\caption{Low angle neutron scattering intensity curves $I(q)$ for a series of
neutral
PDA with increasing density.}
\end{figure}
\begin{figure}
\caption{Experimental dependances of $a/a_0$ (black
symbols)  and
$\xi/a_0$ (open symbols) as a function of $c = \rho/\rho_S$ (log-log plots) for
a basic sample (case a) and for two neutral samples (case b). In case (a) the
full line is the theoretical curve obtained with the off-lattice model with
$c_0=
0.04$ and the open circles are the results obtained with the on-lattice model
with $c_0= 0.08$.}
\end{figure}


\begin{references}


\bibitem{1} W. D. Kingery and M. Berg, J. Appl. Phys., {\bf 26}, 1205, 1955.

\bibitem{2} J. Frenkel, J. Phys. (Moscow), {\bf 9}, 385, 1945.

\bibitem{3} J. K. Mackenzie and R.
Shuttleworth, Proc. Phys. Soc. London, {\bf 62}, 833, 1949.

\bibitem{4} G. W. Scherer, J. Amer. Ceram. Soc, {\bf 60}, 236, 1977; {\it
ibid.}, {\bf
60}, 243, 1977.

\bibitem{5} G. W. Scherer, J. Amer. Ceram. Soc., {\bf 74}, 1523, 1991.

\bibitem{6} R. Semp\'er\'e, D. Bourret, T. Woignier, J. Phalippou and R.
Jullien,
Phys. Rev. Lett., {\bf 71}, 3307, 1993.

\bibitem{7} A. Hasmy, \'E. Anglaret, M. Foret, J. Pelous and R. Jullien, Phys.
Rev.
B, {\bf 50}, 1305, 1994; see also A. Hasmy, \'E. Anglaret,  M.
Foret, J. Pelous, R. Vacher and R. Jullien, in this issue of J. of Non
Cryst. Sol.

\bibitem{8} B. B. Mandelbrot, ``The Fractal Geometry of Nature'', Freeman,
New-York,
1982.


\bibitem{9} P. Meakin, Phys. Rev. Lett., {\bf 51}, 1119, 1983.

\bibitem{10} M. Kolb, R. Botet and R. Jullien, Phys. Rev. Lett., {\bf 51},
1123, 1983.

\bibitem{11} A. Hasmy, M. Foret, J. Pelous and R. Jullien, Phys. Rev. B, {\bf
48},
9345, 1993.

\bibitem{12} S. Ma, ``Modern Theory of Critical Phenomena'', Benjamin,
New-York,
1976.

\bibitem{13} R. Jullien and R. Botet, Aggregation and fractal aggregates, World
Scientific, Singapore, 1987.

\bibitem{14} T. Woignier, habilitation report, University of Montpellier II,
unpublished.

\bibitem{15} A. Hasmy, I. Beurroies, D. Bourret and R. Jullien, preprint.

\end{references}
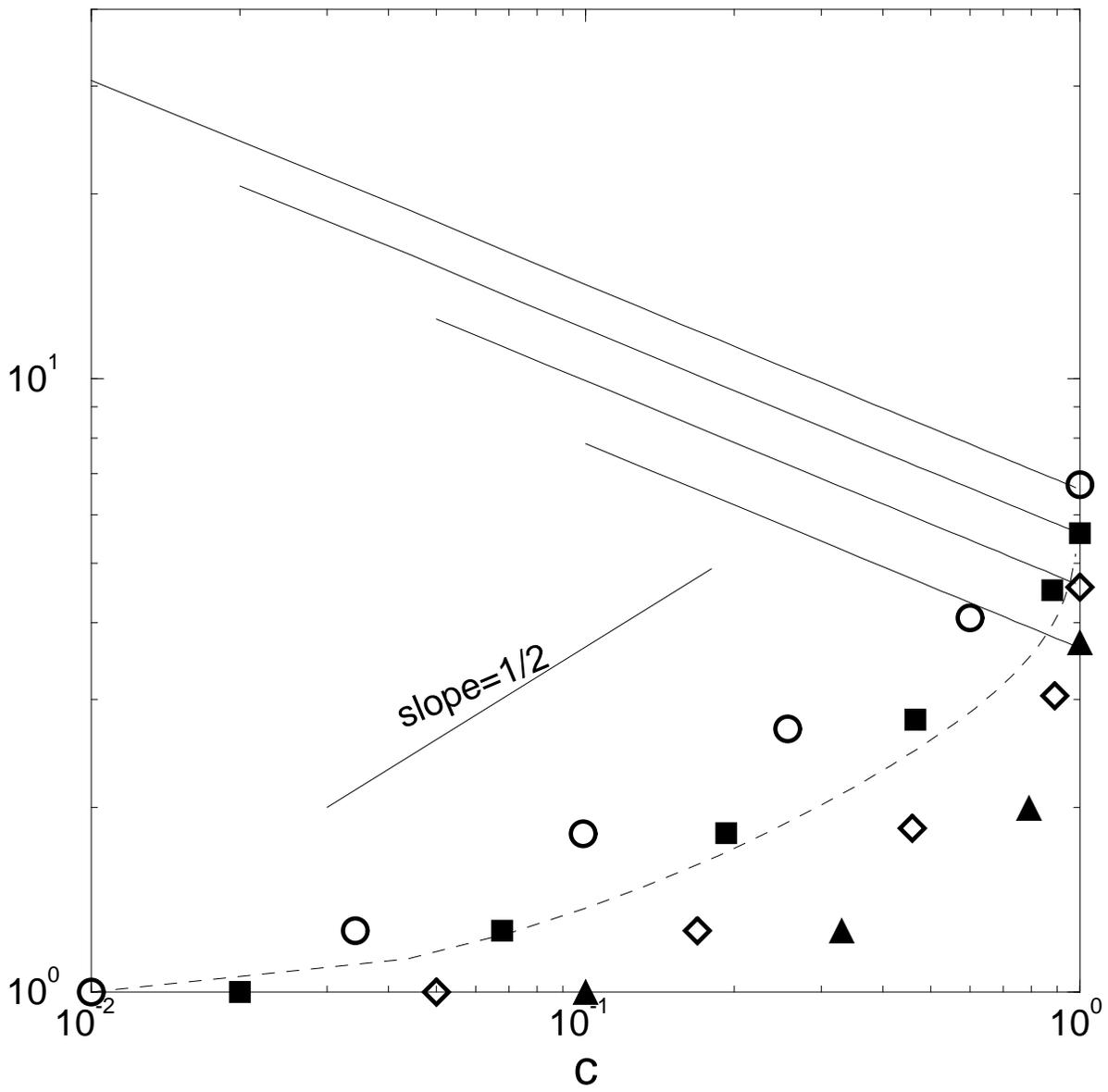
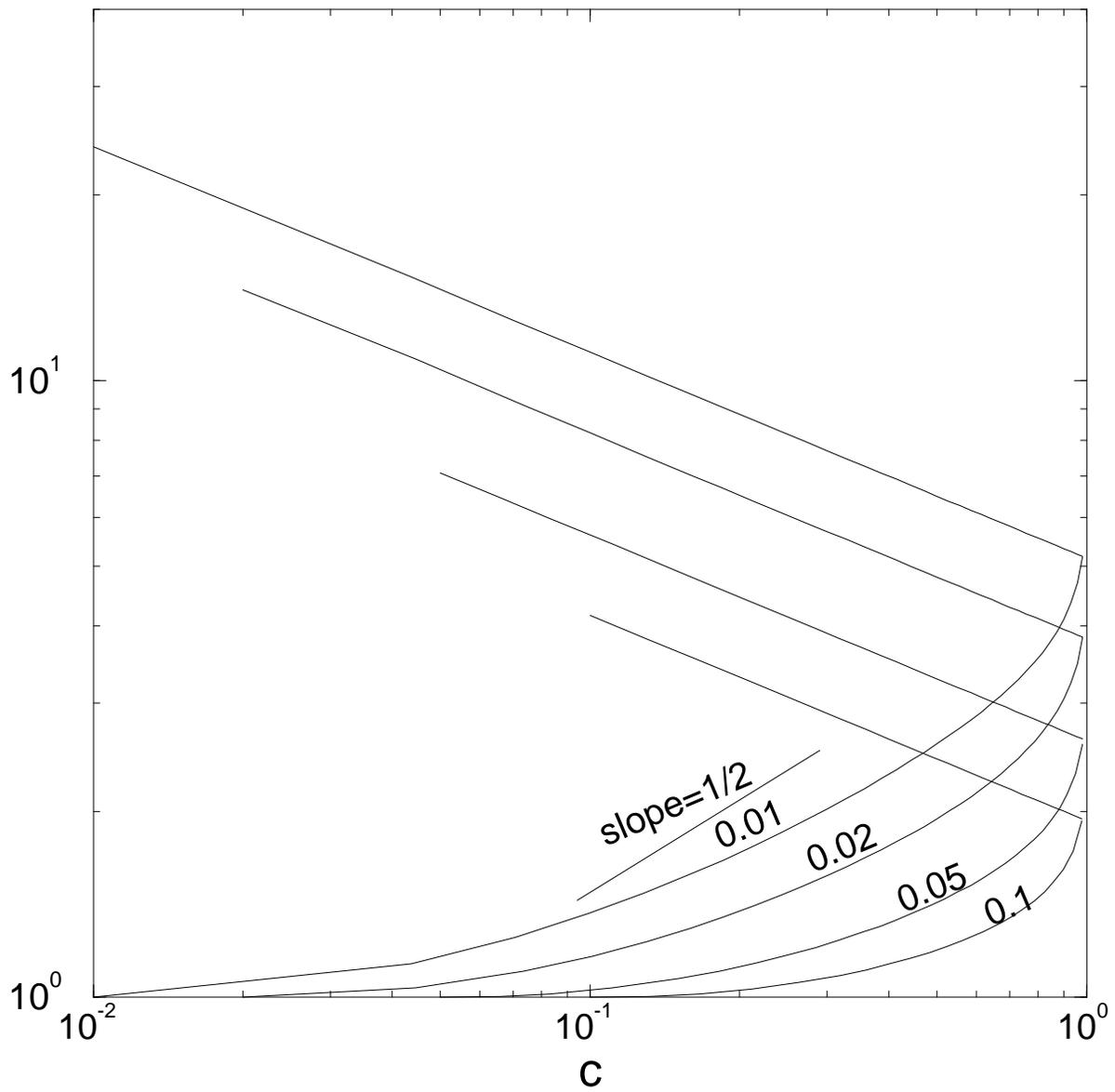
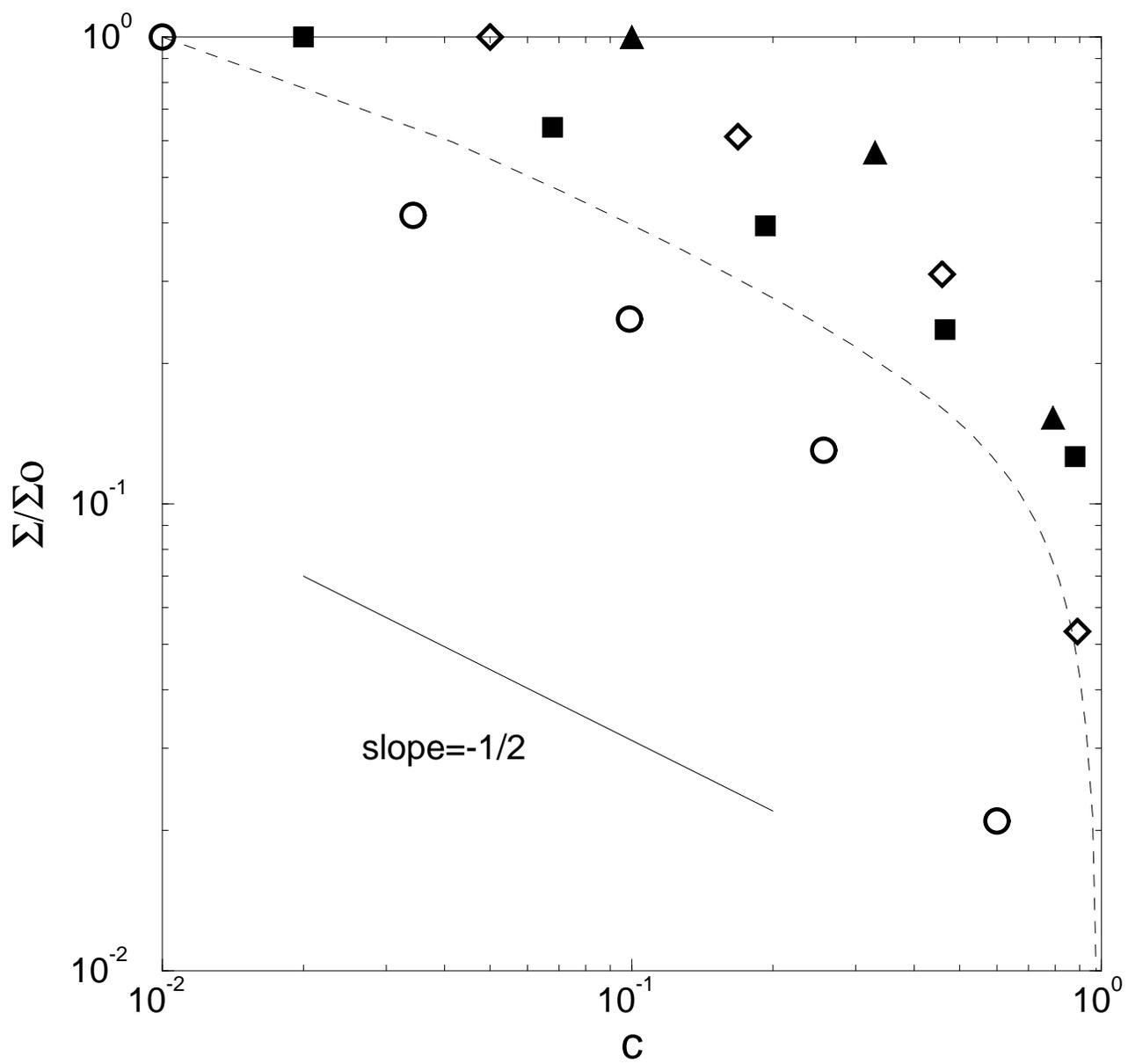
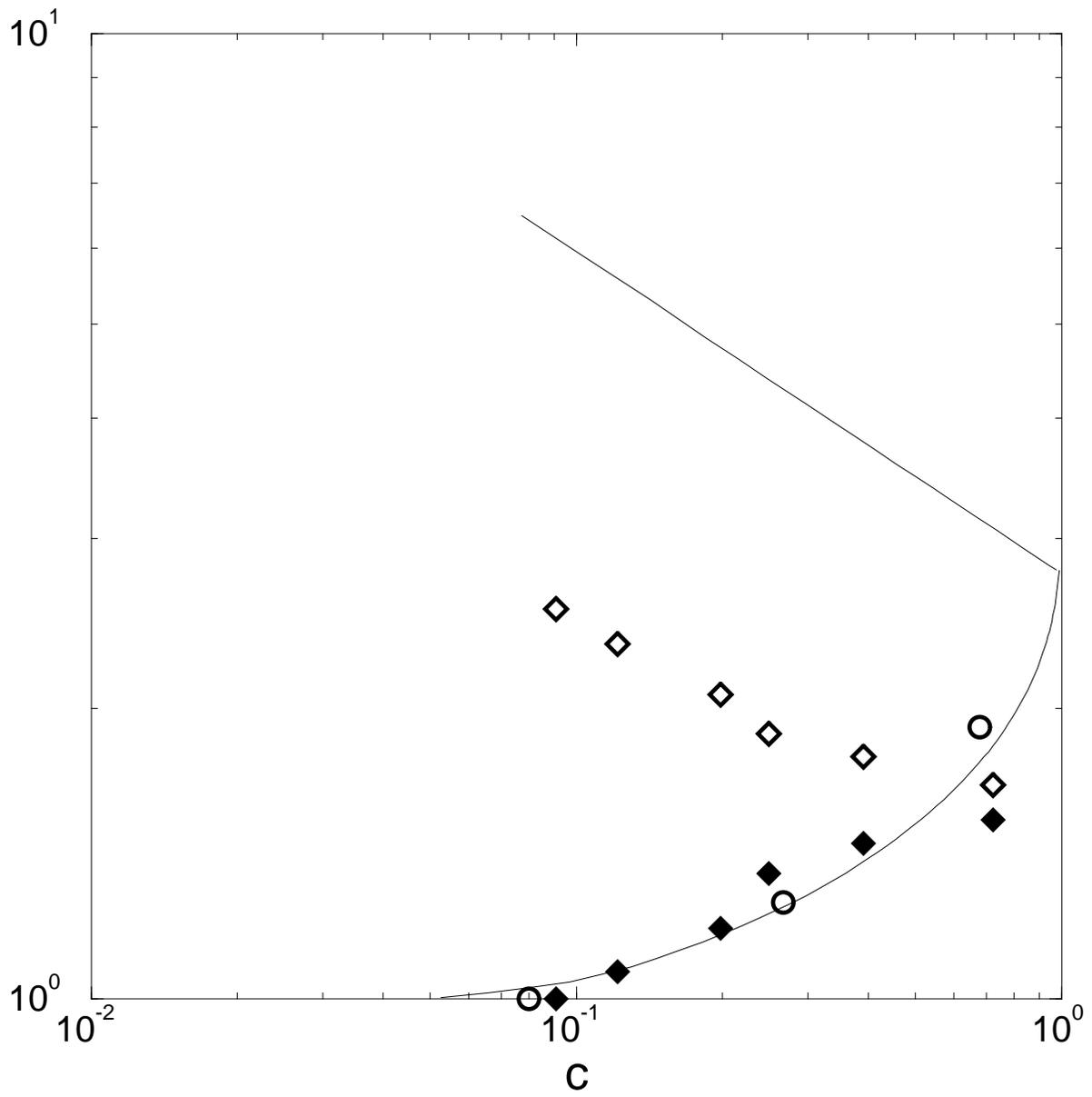
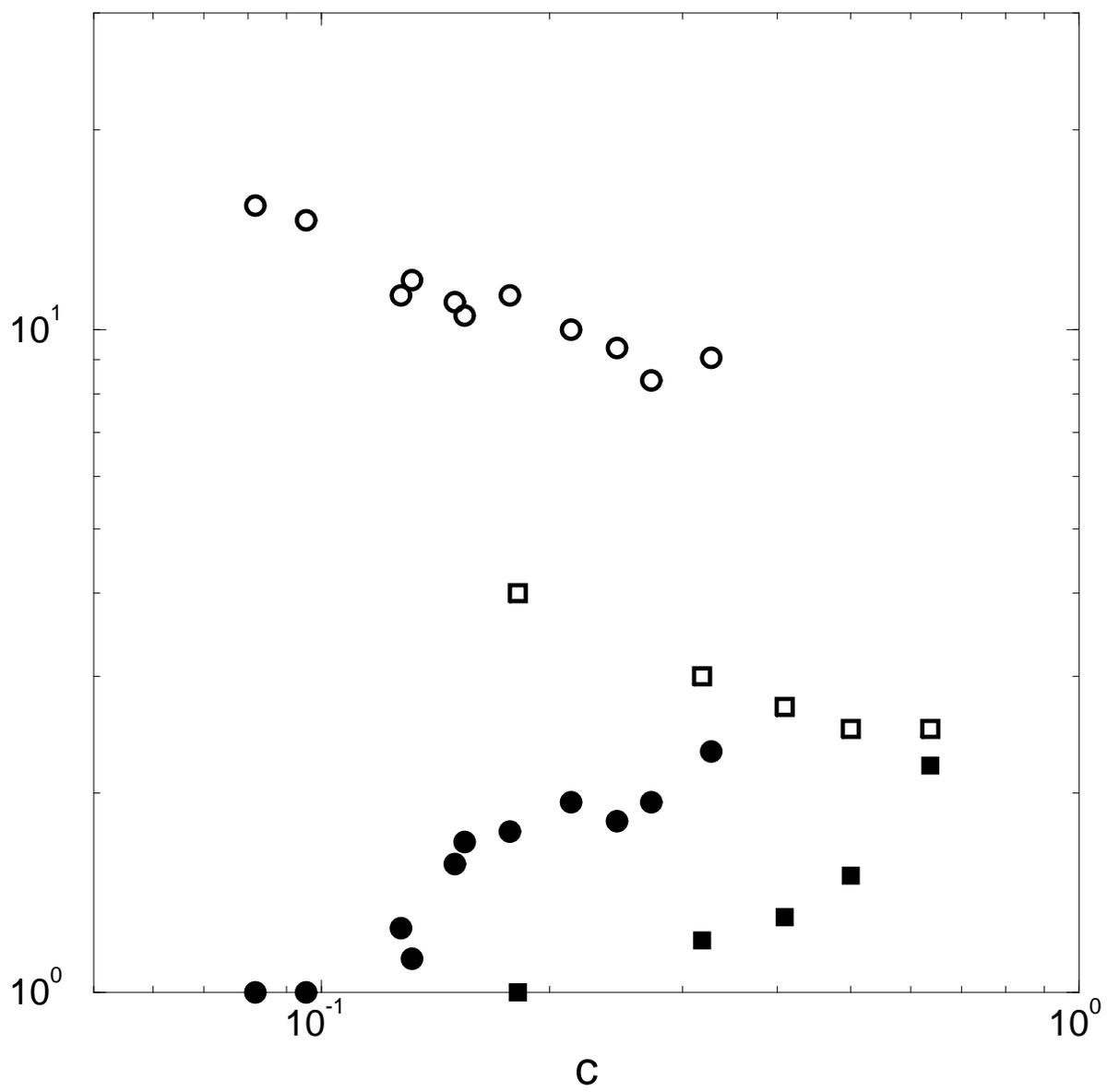
\end{document}